\documentclass[12pt]{article}
\usepackage{graphicx}
\usepackage{bm}
\begin{document}
\newcommand{\cd}{\makebox[0.08cm]{$\cdot$}}
\centerline{\bf Early Thermalization at RHIC}
\vskip 18pt
\centerline{Xiao-Ming X${\rm u}^{{\rm a},{\rm b},{\rm c}}$} 
\vskip 14pt
\centerline{$^{\rm a}$Nuclear Physics Division, Shanghai Institute of Applied 
Physics}
\centerline{Chinese Academy of Sciences, P.O.Box 800204, Shanghai 201800,
            China}
\centerline{$^{\rm b}$Physics Division, Oak Ridge National Laboratory}
\centerline{MS-6373, P.O. Box 2008, Oak Ridge, TN 37831-6373}
\centerline{$^{\rm c}$Department of Physics, Shanghai University, Baoshan, 
Shanghai 200436, China}

\begin{abstract}
\baselineskip=14pt
Triple-gluon elastic scatterings are briefly reviewed since the scatterings
explain the early thermalization puzzle in Au-Au collisions at RHIC energies.
A numerical solution of the transport equation with the triple-gluon elastic 
scatterings demonstrates gluon momentum isotropy achieved at a time of the 
order of 0.65 fm/$c$. 
Triple-gluon scatterings lead to a short thermalization time of gluon matter.  
\end{abstract}
\leftline{Keywords: Triple-gluon elastic scatterings; transport equation; 
thermalization}

\vspace{0.5cm}
\leftline{\bf 1. Introduction}
\vspace{0.5cm}
Hadron spectra at low momentum can be fitted to thermal distributions. The 
disappearance of angular correlation of back-to-back jets means that one jet 
is 
dissolved into medium and thermalized. The thermalization of matter created
in Au-Au collisions at RHIC is obvious. However, relevant importance is 
that the initially created matter
thermalizes within a time of less than 1 fm/$c$ [1,2] at least in the regime
near midrapidity [3]. The early thermalization is concluded while the
results of hydrodynamic calculations [1,3-7] are in agreement with
the elliptic flow data [8,9] at $p_{\bot} < 2$ GeV/$c$. Triple-gluon elastic 
scatterings were proposed to explain the early thermalization puzzle in Ref.
[10].   

Why the gluon-gluon-gluon elastic scattering processes are needed at RHIC 
energies? 
The three-gluon to three-gluon scatterings get important
while the gluon number density is high. To account for RHIC experimental data 
the gluon rapidity density
$dN^{\rm g}/dy \sim 1000$ is suggested for the initial gluon matter [11,12]. 
Such a value of the rapidity density leads to 
a gluon number density of about 38 ${\rm fm}^{-3}$, 
which is high enough for triple-gluon scatterings to occur. 
When the gluon number density gets larger, three-gluon scatterings 
become more important. The three-gluon scattering processes are anticipated
to thoroughly overwhelm the two-gluon
elastic scattering processes in heavy ion collisions at LHC energies. 
An estimate from minijet production gives initial gluon number densities 
of about 30 ${\rm fm}^{-3}$ for RHIC and 140 ${\rm fm}^{-3}$ for LHC [13].
This supports the importance of the three-gluon scattering processes.

\begin{center}
      {\includegraphics[width=35mm,height=55mm,angle=0]{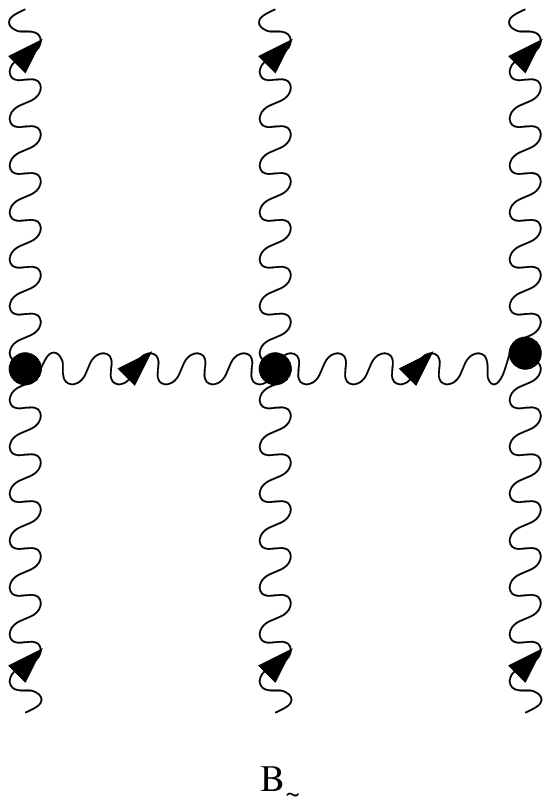}}
       \hspace{1cm} 
      {\includegraphics[width=35mm,height=55mm,angle=0]{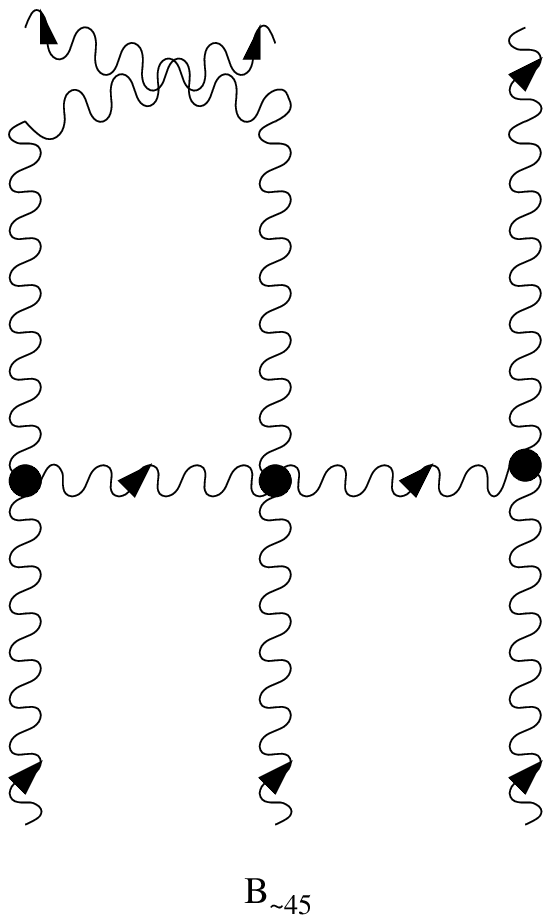}}
       \hspace{1cm} 
      {\includegraphics[width=35mm,height=55mm,angle=0]{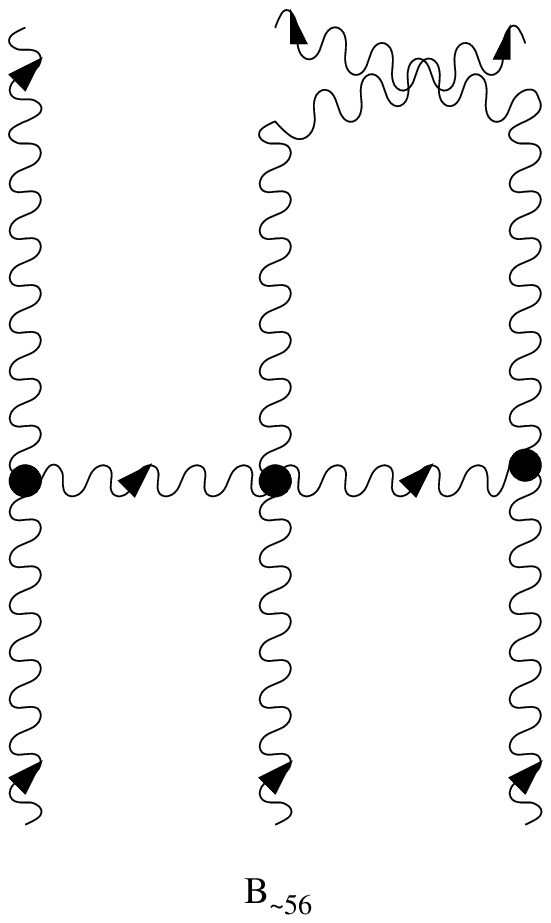}}
       \vskip 26pt
      {\includegraphics[width=35mm,height=55mm,angle=0]{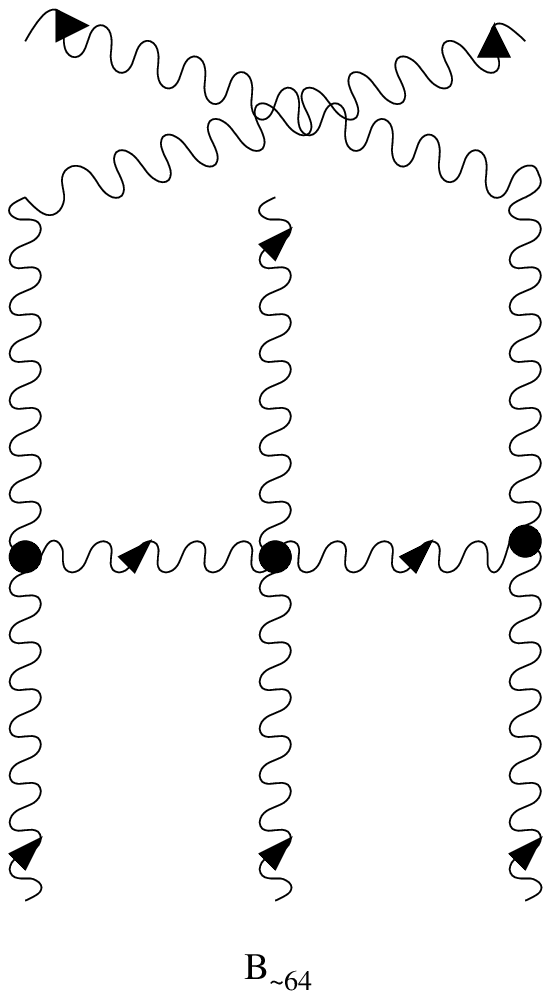}}
       \hspace{1cm}   
      {\includegraphics[width=35mm,height=55mm,angle=0]{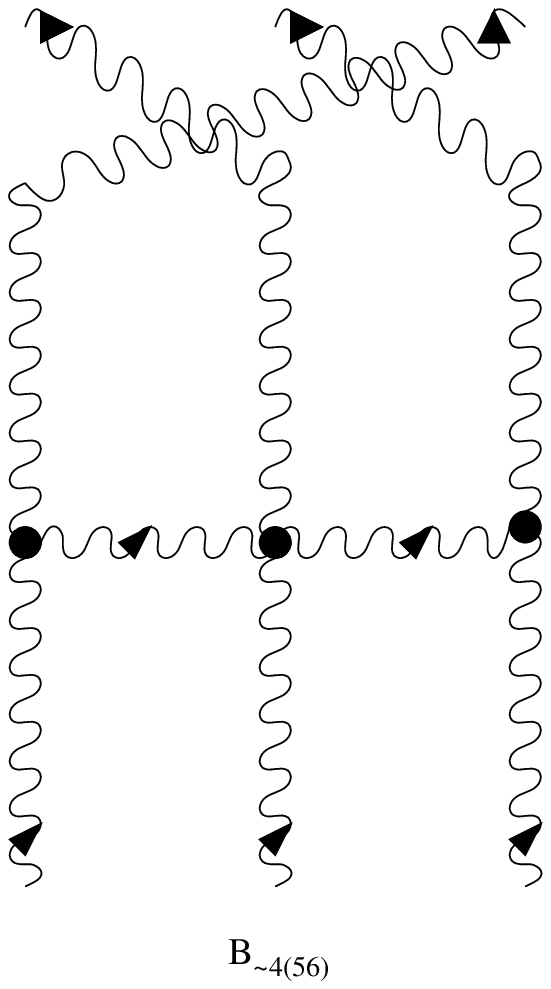}}
       \hspace{1cm}   
      {\includegraphics[width=35mm,height=55mm,angle=0]{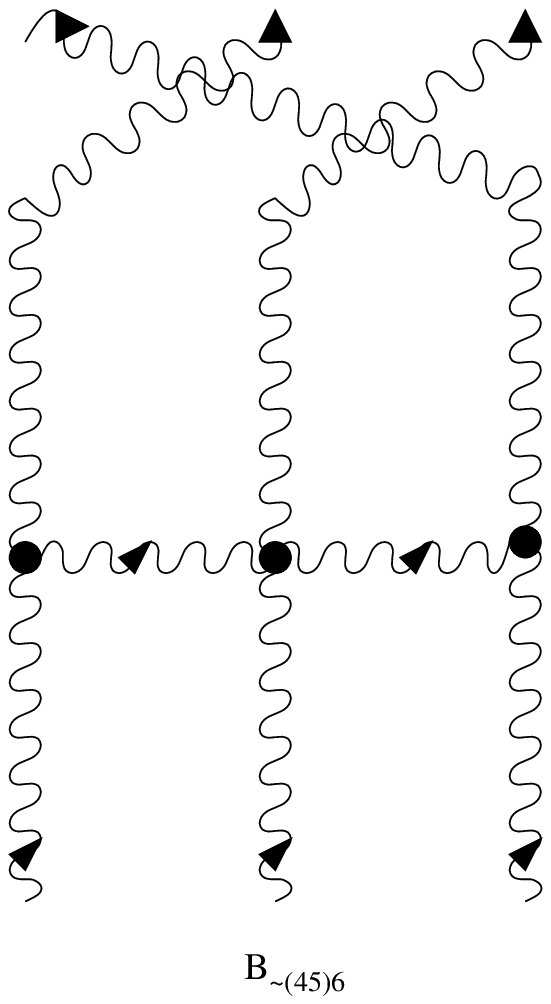}}
\end{center}
\leftline{Fig. 1: Scatterings of three gluons.}

\vspace{0.5cm}
\leftline{\bf 2. Triple-gluon scatterings}
\vspace{0.5cm}
The two-gluon to two-gluon scattering processes  
were studied in perturbative QCD by Cutler and Sivers [14], 
Combridge, Kripfganz and Ranft [15]. 
The spin- and color-averaged squared amplitude
$\mid {\cal M}_{2 \to 2} \mid^2$ for the two-gluon elastic scatterings 
at order $\alpha_{\rm s}^2$ was  
expressd in terms of the Mandelstam variables.
For the triple-gluon elastic scatterings, many diagrams have to be calculated  
by fortran codes. 
Some of the three-gluon to three-gluon scattering diagrams are presented 
in Figs. 1 and 2
and corresponding spin- and color-averaged squared amplitude 
$\mid {\cal M}_{3 \to 3} \mid^2$ is calculated in perturbative QCD.
A gluon four-momentum is labeled as 
$p_{\rm i}=(E_{\rm i},\vec {p}_{\rm i})$ in the process
${\rm g}(p_1)+{\rm g}(p_2)+{\rm g}(p_3) \to
{\rm g}(p_4)+{\rm g}(p_5)+{\rm g}(p_6)$.
Then nine Lorentz-invariant variables are defined as
\begin{displaymath}
s_{12}=(p_1+p_2)^2,~~~~~~~~s_{23}=(p_2+p_3)^2,~~~~~~~~s_{31}=(p_3+p_1)^2
\end{displaymath}
\begin{displaymath}
u_{15}=(p_1-p_5)^2,~~~~~~~~~~~~~~~~~~u_{16}=(p_1-p_6)^2
\end{displaymath}
\begin{displaymath}
u_{24}=(p_2-p_4)^2,~~~~~~~~~~~~~~~~~~u_{26}=(p_2-p_6)^2
\end{displaymath}
\begin{displaymath}
u_{34}=(p_3-p_4)^2,~~~~~~~~~~~~~~~~~~u_{35}=(p_3-p_5)^2
\end{displaymath}
These independent variables are  
appropriate for expressing $\mid {\cal M}_{3 \to 3} \mid^2$. 

\begin{center}
       \leavevmode
      {\includegraphics[width=35mm,height=45mm,angle=0]{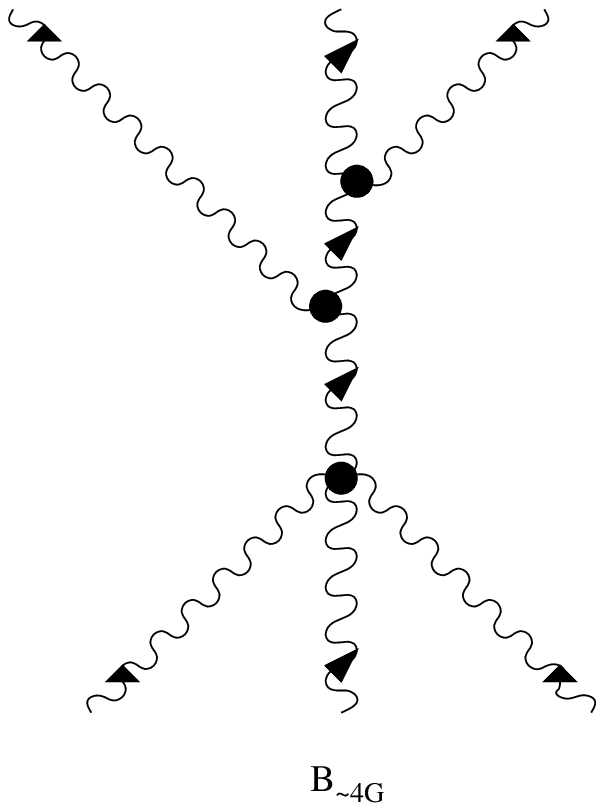}}
       \hspace{1cm}   
      {\includegraphics[width=35mm,height=45mm,angle=0]{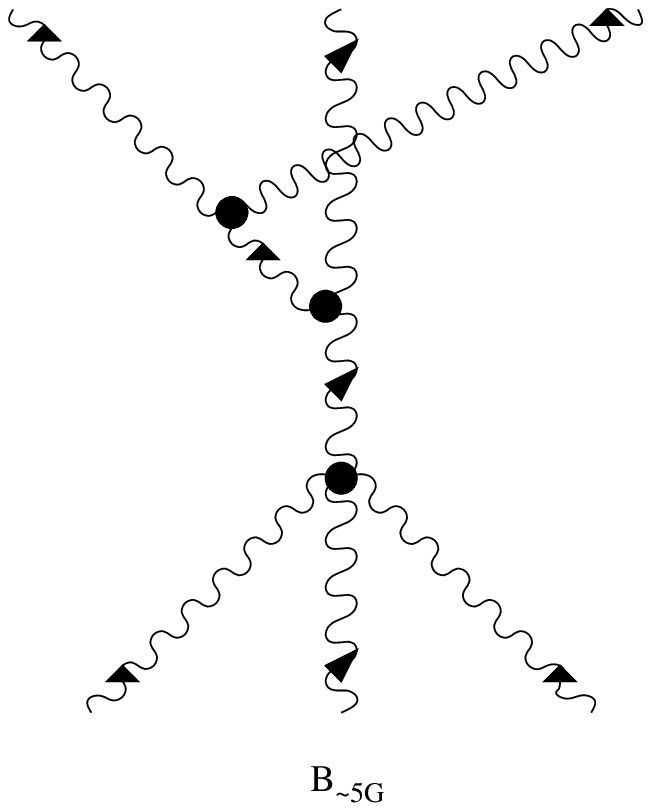}}
       \hspace{1cm} 
      {\includegraphics[width=35mm,height=45mm,angle=0]{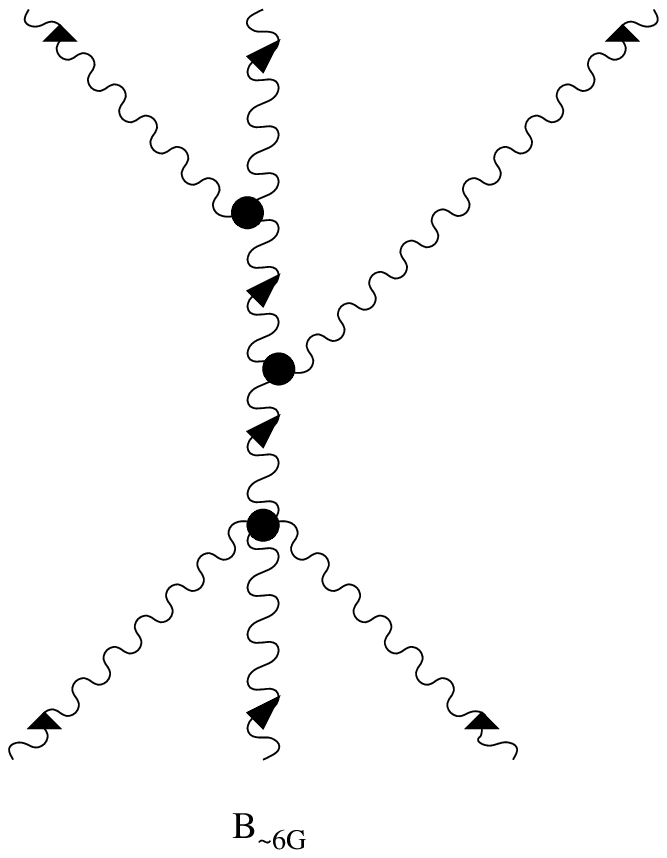}}
       \vskip 26pt
      {\includegraphics[width=45mm,height=45mm,angle=0]{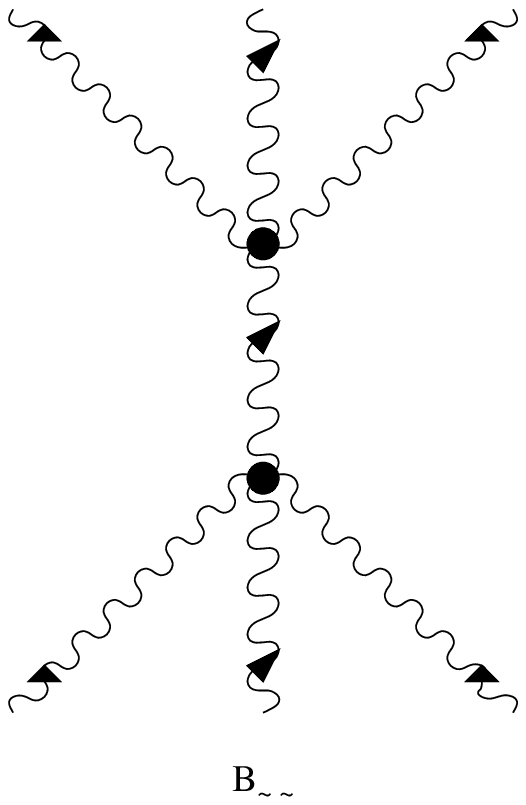}}
\end{center}
\leftline{Fig. 2: Scatterings of three gluons.} 

\vspace{0.5cm}
Only the diagram ${\rm B}_{\sim \sim}$ in Fig. 2 is calculable by 
hand. This process consists of two four-gluon couplings and one 
gluon propagator between the two vertices.
The spin- and color-averaged squared amplitude 
$\mid  {\cal M}_{{\rm B}_{\sim \sim}} \mid^2$ for the diagram 
${\rm B}_{\sim \sim}$ then depends on the  three variables
$s_{12}$, $s_{23}$ and $s_{31}$ in the simple form
\begin{equation}
\mid  {\cal M}_{{\rm B}_{\sim \sim}} \mid^2 
=\frac {59049}{128}\frac {{\rm g}_{\rm s}^8} 
{(s_{12}+s_{23}+s_{31})^2}
\end{equation}
which is independent 
of the momenta of the three final gluons. This ensures that
each of the final gluons runs away in any direction 
in momentum space  with equal probability.
In other words, local momentum isotropy must be attained through the scattering
process. Here, ${\rm g}_{\rm s}$ is the quark-gluon coupling constant,
${\rm g}_{\rm s}^2 =4\pi \alpha_{\rm s}$ and $\alpha_{\rm s}=0.3$ is used

\vspace{0.5cm}
\leftline{\bf 3. Thermalization of gluon matter}
\vspace{0.5cm}
Since the triple-gluon elastic scatterings are important in the initial gluon
matter, a transport equation of Boltzmann type must include the $3 \to 3$
scattering term,
\begin{eqnarray}
& & 
\frac {\partial f_1}{\partial t} 
+ \vec {\rm v}_1 \cdot \vec {\nabla}_{\vec r} f_1 
         \nonumber    \\
& &
= -\frac {{\rm g}_{\rm G}}{2E_1{\rm g}_{22}} \int \frac {d^3p_2}{(2\pi)^32E_2}
\frac {d^3p_3}{(2\pi)^32E_3} \frac {d^3p_4}{(2\pi)^32E_4}  
(2\pi)^4 \delta^4(p_1+p_2-p_3-p_4)
         \nonumber    \\
& &
~~~ \times \mid {\cal M}_{2 \to 2} \mid^2
[f_1f_2(1+f_3)(1+f_4)-f_3f_4(1+f_1)(1+f_2)]
         \nonumber    \\
& &
~~~ -\frac {{\rm g}_{\rm G}^2}{2E_1{\rm g}_{33}} 
\int \frac {d^3p_2}{(2\pi)^32E_2}
\frac {d^3p_3}{(2\pi)^32E_3} \frac {d^3p_4}{(2\pi)^32E_4}  
\frac {d^3p_5}{(2\pi)^32E_5} \frac {d^3p_6}{(2\pi)^32E_6}  
         \nonumber    \\
& &
~~~ \times (2\pi)^4 \delta^4(p_1+p_2+p_3-p_4-p_5-p_6)
\mid {\cal M}_{3 \to 3} \mid^2 
         \nonumber    \\
& &
~~~ \times [f_1f_2f_3(1+f_4)(1+f_5)(1+f_6)-f_4f_5f_6(1+f_1)(1+f_2)(1+f_3)]
         \nonumber    \\
\end{eqnarray}
where ${\rm g}_{22}=2$, ${\rm g}_{33}=12$, 
the degeneracy factor ${\rm g}_{\rm G}=16$
and the velocity of a massless gluon ${\rm v}_1=1$.
The gluon distribution function $f_{\rm i}$ depends on 
the position $\vec {r}_{\rm i}$, the momentum $\vec {p}_{\rm i}$ 
and the time $t$. The first term on the right-hand-side of the equation 
exhibits the well-known $2 \to 2$ scattering processes.
The second term is a new term, which 
represents the $3 \to 3$ scatterings. Six gluon distributions are involved
in the new term. The $3 \to 3$ scattering processes involve a larger phase 
space than the $2 \to 2$ scattering processes.

The squared amplitude $\mid {\cal M}_{3 \to 3} \mid^2$ has a very much long
expression and substantially enhance the running time of computer in solving 
the transport equation.  
To gain a first sight at the physics determined by the transport equation,
initially produced gluons are assumed to uniformly 
distribute in a cylinder formed in a central heavy ion collision.
Consequently, the gluon distribution function is only a function of momentum 
magnitude and time. 
We solve the equation starting at the time $t_{\rm ini}$ when  gluon matter
is created and ending at the time $t_{\rm iso}$ when gluon matter reaches local
momentum isotropy. The transverse expansion of the initially produced 
gluon matter is neglected due to a short thermalization time.

\begin{center}
      {\includegraphics[width=0.7\textwidth,angle=0]{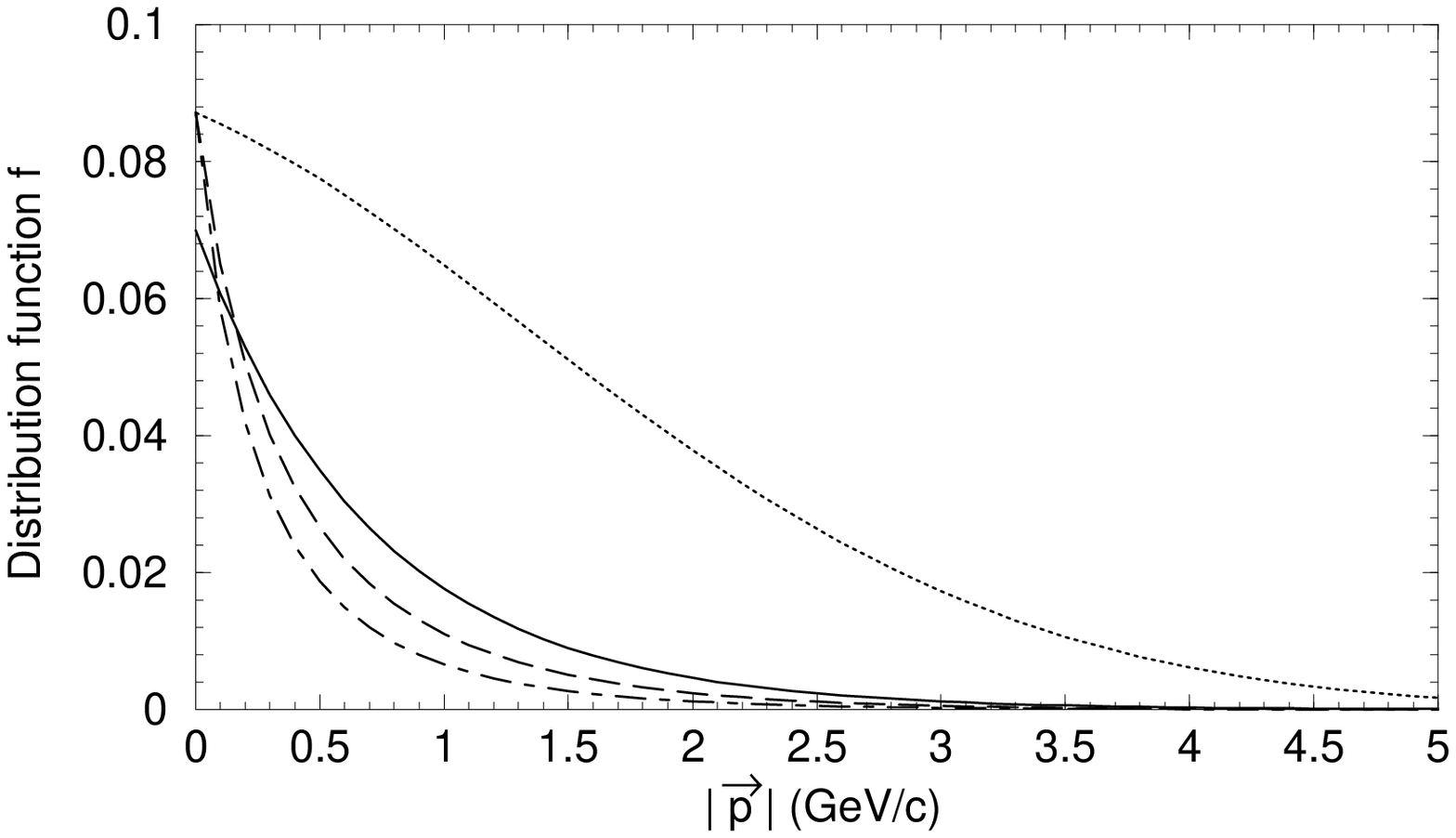}}
\end{center}
\noindent{Fig. 3: Gluon distribution functions versus momentum in different
directions while  gluon matter is just produced in 
the initial Au-Au collision.
The dotted, dashed and dot-dashed curves correspond to the angles 
$\theta =0^{\rm o}, 45^{\rm o}, 90^{\rm o}$, 
respectively. The solid curve stands for the 
thermal distribution function.}

\vspace{0.5cm} 
For a central Au-Au collision at $\sqrt {s_{NN}}=200$ GeV,
an initial gluon distribution for the transport
equation is obtained from Eq. (34) of Ref. [16],
\begin{equation}
f(\vec {p},t_{\rm ini})=\frac {1.71\times 10^7 (2\pi)^{1.5}}
{{\rm g}_{\rm G}\pi R_A^2 Y(\mid \vec {p} \mid/\cosh ({\rm y})+0.3)}
{\rm e}^{-\mid \vec {p} \mid/(0.9\cosh ({\rm y}))-(\mid \vec {p} \mid 
\tanh ({\rm y}))^2/8} \bar {\theta} (Y^2-{\rm y}^2)
\end{equation}
where $\mid \vec {p} \mid$ is in GeV,
the nuclear radius $R_A=6.4$ fm and y is the rapidity. 
The time $t_{\rm ini}$ is 
0.2 fm/$c$ as estimated in  HIJING Monte Carlo simulation [17]. Here,
$Y$ is the maximum of rapidity and approximately equals 5. 
$\bar {\theta} (Y^2-{\rm y}^2)$ is the step function which is zero for $Y<y$
or 1 for $Y \ge y$.
Eq. (3) exhibits one form of initial gluon distribution in  gluon matter
which is not in thermal and chemical equilibrium. We draw in Fig. 3 the 
dotted, dashed and dot-dashed curves individually for the gluon 
distribution functions against the gluon momentum at the three angles 
$\theta = 0^{\rm o}, 45^{\rm o}, 90^{\rm o}$ relative to the incoming beam
direction. These curves can not coincide and do not 
exhibit local momentum isotropy. Particularly, the gluon momentum
distribution in the transverse direction differs considerably from one in the
longitudinal direction.

\begin{center}
      {\includegraphics[width=0.7\textwidth,angle=0]{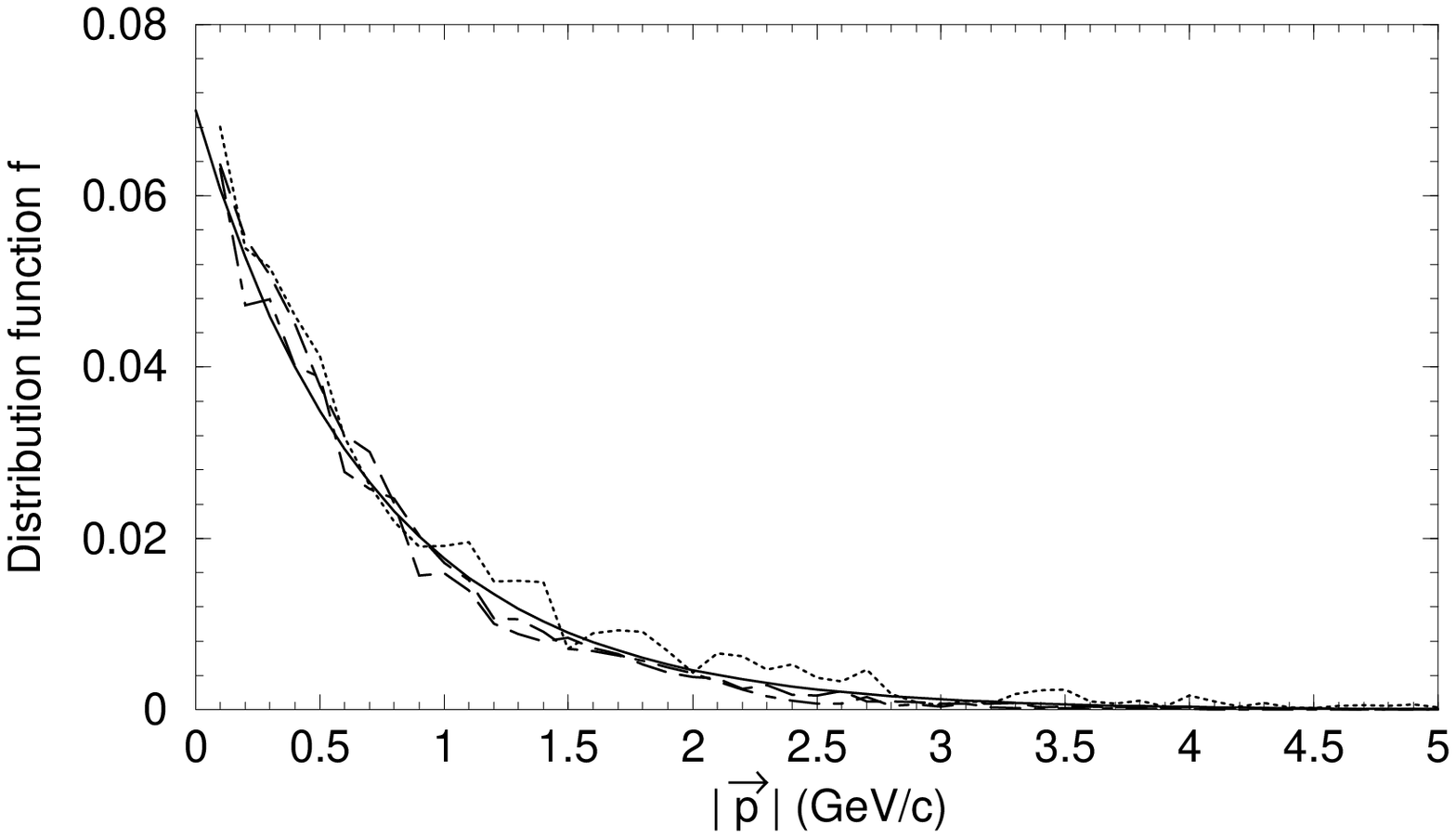}}
\end{center}
\noindent{Fig. 4: Gluon distribution functions versus momentum in different
directions while  gluon matter just arrives at thermal equilibrium.
The dotted, dashed and dot-dashed curves correspond to the angles  
$\theta =0^{\rm o}, 45^{\rm o}, 90^{\rm o}$, respectively. 
The solid curve represents the 
thermal distribution function.}

\vspace{0.5cm}
The difference among the initial gluon distribution functions in different 
directions does not last for a long time since the transport equation alters 
them. Momentum isotropy can be established at the time $t_{\rm iso}=0.65$ 
fm/$c$ as shown by the dotted, dashed and dot-dashed curves in Fig. 4.   
The gluon distribution functions can be well fitted to the J$\rm \ddot u$ttner 
distribution with nonequilibrium fugacity $\lambda$,
\begin{equation}
f(\vec {p},t_{\rm iso})=\frac {\lambda}{{\rm e}^{\mid \vec {p} \mid/T}-\lambda}
\end{equation}
where $T$ gives the temperature of gluon matter. The solid curve in Fig. 4 
gives a fit of $\lambda=0.065$ and $T=0.75$ GeV.
We get the thermalization time $t_{\rm iso} - t_{\rm ini}=0.45$ fm/$c$.
The fugacity and the temperature are higher than 0.05 and 0.55 GeV, and
the thermalization time is less than 0.5 fm/$c$ 
obtained from the free streaming of partons in Ref. [16], respectively.

\vspace{0.5cm}
\leftline{\bf 4. Summary}
\vspace{0.5cm}
The three-gluon scattering processes proposed to study 
the early thermalization problem at high density have been reviewed. 
The gauge-invariant squared amplitude
$\mid {\cal M}_{3 \to 3} \mid^2$ for all the triple-gluon scatterings
is given at the lowest order $\alpha_{\rm s}^4$.
$\mid {\cal M}_{3 \to 3} \mid^2$ enters the transport equation to give a new
contribution to the time dependence of the gluon distribution function. 
The evolution of gluon matter is dominated by 
the three-gluon scattering processes while the number density is high. 
With the initial gluon matter
obtained from  HIJING simulation, the transport equation gives the 
thermalization time of about 0.45 fm/$c$. 
The three-gluon scatterings considerably shorten the thermalization time of 
gluon matter at high density.

\vspace{0.5cm}
\leftline{\bf Acknowledgements}
\vspace{0.5cm}
X.M. Xu thanks Xin-Nian Wang and Wei Zhu for comments.
This work was supported in part by National Natural Science Foundation of China
under Grant No. 10135030, in part by Shanghai Education Committee Research 
Fund, in part by the CAS
Knowledge Innovation Project No. KJCX2-SW-N02, in part by the Division of
Nuclear Physics, Department of Energy, under Contract No. DE-AC05-00OR22725 
managed by UT-Battelle, LLC.

\vspace{0.5cm}
\leftline{\bf References}
\vskip 14pt
\leftline{[1]U. Heinz, P.F. Kolb, in: R. Bellwied, J. Harris, W. Bauer (Eds.), 
Proc. of}
\leftline{~~~the 18th Winter Workshop on Nuclear
Dynamics, EP Systema, Debrecen,}
\leftline{~~~Hungary, 2002.}
\leftline{[2]E.V. Shuryak, Nucl. Phys. A715(2003)289c.}  
\leftline{[3]T. Hirano, Phys. Rev. C65(2001)011901.}
\leftline{[4]P. Huovinen, Nucl. Phys. A715(2003)299c.}
\leftline{[5]K. Morita, S. Muroya, C. Nonaka, T. Hirano, Phys. Rev. 
C66(2002)054904.}
\leftline{[6]D. Teaney, J. Lauret, E.V. Shuryak, nucl-th/0110037.}
\leftline{[7]K.J. Eskola, et al., Phys. Lett. B566(2003)187;}
\leftline{~~~K.J. Eskola, et al., Nucl. Phys. A715(2003)561c.}
\leftline{[8]K.H. Ackermann, et al., STAR Collaboration, Phys. Rev. Lett. 
86(2001)402;}
\leftline{~~~R.J. Snellings, et al., for the STAR Collaboration, Nucl. Phys. 
A698(2002)193c;} 
\leftline{~~~C. Adler, et al., STAR Collaboration, Phys. Rev. Lett. 
87(2001)182301;}
\leftline{~~~C. Adler, et al., STAR Collaboration, Phys. Rev. 
C66(2002)034904;} 
\leftline{~~~C. Adler, et al., STAR Collaboration, Phys. Rev. Lett. 
90(2003)032301.}
\leftline{[9]R.A. Lacey, et al., for the PHENIX Collaboration, Nucl. Phys. 
A698(2002)559c;}
\leftline{~~~S.S. Adler, et al., PHENIX Collaboration, Phys. Rev. Lett. 
91(2003)182301.} 
\leftline{[10]X.-M. Xu, Y. Sun, A.-Q. Chen, L. Zheng, Nucl. Phys. 
A744(2004)347.}
\leftline{[11]K.J. Eskola, K. Kajantie, K. Tuominen, Phys. Lett. B497(2001)39.}
\leftline{[12]M. Gyulassy, P. L$\rm \acute e$vai, I. Vitev, 
Nucl. Phys. B594(2001)371;} 
\leftline{~~~~M. Gyulassy, I. Vitev, X.-N. Wang, P. Huovinen, 
Phys. Lett. B526(2001)301.} 
\leftline{[13]F. Cooper, E. Mottola, G.C. Nayak, Phys. Lett. B555(2003)181.}
\leftline{[14]R. Cutler, D. Sivers, Phys. Rev. D17(1978)196.}
\leftline{[15]B.L. Combridge, J. Kripfganz, J. Ranft, Phys. Lett. 
70B(1977)234.}
\leftline{[16]P. L$\rm \acute e$vai, B. M$\rm \ddot u$ller, X.-N. Wang,
Phys. Rev. C51(1995)3326.} 
\leftline{[17]X.-N. Wang, M. Gyulassy, Phys. Rev. D44(1991)3501;}
\leftline{~~~~X.-N. Wang, M. Gyulassy, Comput. Phys. Commun. 83(1994)307;}
\leftline{~~~~X.-N. Wang, Phys. Rep. 280(1997)287.}

\end{document}